\documentstyle[aps,preprint]{revtex}
\begin{document}
\draft
%\title{Can Thermal Noise Facilitate Energy Conversion 
%by Ratchet System?}
\title{Thermal noise can facilitate energy conversion by a ratchet system}
\author{Fumiko Takagi\footnote{Electronic address:
fumiko@cmpt01.phys.tohoku.ac.jp} and 
Tsuyoshi Hondou\footnote{Electronic address: hondou@cmpt01.phys.tohoku.ac.jp}}
\address{Department of Physics, Tohoku University 
        \\ Sendai 980-8578, Japan}
\date{Received 26 March 1999}
\maketitle
\begin{abstract}
Molecular motors in biological systems
are expected to use ambient fluctuation.
In a recent Letter
[Phys. Rev. Lett. {\bf 80}, 5251 (1998)],
it was showed
that the following question was unsolved,
``Can thermal noise facilitate energy conversion by ratchet system?''
We consider it using stochastic energetics,
and show that there exist systems where thermal noise
helps the energy conversion.
\end{abstract}

%--------- Introduction ----------
Molecular motors in biological systems are known to operate 
efficiently\cite{Yanagida_ea-Slidi,Ishijima-Multi,Uyeda_ea-Myosi,Yasuda_ea}.
They convert molecular scale chemical energy 
into macroscopic mechanical work
with high efficiency in water at room temperature,
where the effect of
thermal fluctuation is unavoidable.
These experimental facts % about molecular motors
lead us to expect the existence of the system where thermal 
noise helps the operation.
To find out the mechanism of these motors
is interesting 
not only to biology
but also to statistical and thermal physics.

%--------
Recently inspired by observations on the molecular motors,
many studies have been performed 
from the viewpoint of statistical physics.
Much has been studied in ratchet
models\cite{Vale_ea-Prote,AHuxley,Julicher-Model}
to consider how the directed motion appears 
from non-equilibrium fluctuation.
%---Magnasco
One of the best known works among these ratchet models
was by Magnasco\cite{Magnasco-Force}.
He studied ``forced thermal ratchet,''
and claimed that ``there is a region of the operating regime
where the efficiency is optimized at finite temperatures.''
His claim is interesting because 
thermal noise is usually known
to disturb the operation of machines.
%---Kamegawa, et al.
However, recently
it was revealed 
that this claim was made incorrectly\cite{Kamegawa_ea-Energ},
because it was not based on the analysis of the energetic efficiency
but only on that of the probability current,
as most of the studies of ratchet systems were.
The insufficient analysis was attributed to 
the lack of systematic method
of energetics in systems described by Langevin equation.
Recently a method what is called stochastic energetics was formalized,
where 
the heat was described quantitatively
in the frame of Langevin equation\cite{Sekimoto-Energ}.
Using this method,
some attempts to discuss the energetics of these
systems\cite{Sekimoto_ea-Compl,Matsuo-Stoch,Hondou_ea-Irrev,Sekimoto_ea-Molec}
have been made. 
By the energetic formulation of the forced thermal 
ratchet\cite{Kamegawa_ea-Energ}
using this stochastic energetics,
the following was showed:
The behavior of the probability current is
qualitatively different than that of energetic efficiency.
Thermal noise does {\it not\/} help 
the energy conversion by the ratchet
at least on the condition
where the claim was made.

Therefore 
it was revealed that
the following question had not yet been solved,
``Can thermal noise facilitate operation of the ratchet?''
In this Letter, we will show that 
the thermal noise certainly can facilitate the operation of the ratchet.

%--------- Oscillating Ratchet ----------
Let us consider an over-dumped particle 
in an ``oscillating ratchet'',
where the amplitude of the 1-D ratchet potential is constant, 
but the degree of the symmetry breaking oscillates at frequency $\omega$
(Fig. \ref{Fig:OscillatingPotential}).
Langevin equation is as follows:
\begin{eqnarray}
 \frac{d x}{d t} &=& 
  -\frac{\partial V(x,t)}{\partial x} + \xi(t),
  \label{LangevinEq} \\
 V(x,t) &=& V_{p}(x,t)+ \ell x,
  \label{OR:potential}
\end{eqnarray}
where $x$, $\ell$ and $V_{p}(x,t)$ represent the state of the system,
the load and the ratchet potential 
respectively (Fig. \ref{Fig:OR:loaded}) .
The white and Gaussian random force $\xi(t)$ satisfies
$\left\langle \xi(t) \right\rangle=0$ and 
$\left\langle \xi(t)\xi(t') \right\rangle=2 \epsilon \delta(t-t')$,
where the angular bracket $\left\langle \cdot \right\rangle$ denotes
the ensemble average.
We use the unit $m=\gamma=1$.
We assume
that the potential $V(x,t)$ always has basins,
and thus a particle cannot move over the potential peak
without thermal noise. % (Fig. \ref{Fig:OR:loaded}).
The ratchet $V_{p}(x,t)$ is assumed to satisfy
the temporally and spatially periodic conditions,
\begin{eqnarray}
 V_{p}(x,t+T) &=& V_{p}(x,t), \label{OR:temp_p}\\
 V_{p}(x+L,t) &=& V_{p}(x,t), \label{OR:spac_p}
\end{eqnarray}  
where $L$ is a spatial period of the ratchet potential, 
and $T \left(\equiv\frac{2 \pi}{\omega}\right)$ 
is a temporal period of the potential modulation.
By potential modulation,
energy is introduced into the system,
and the system converts it into work against the load\cite{fn:chemical}.

The Fokker-Planck equation\cite{Risken_FPbook}
corresponding to Eq. (\ref{LangevinEq})
is written
\begin{eqnarray}
 \frac{\partial P(x,t)}{\partial t}
  &=& - \frac{\partial J(x,t)}{\partial x},\nonumber\\
  &=& - \frac{\partial}{\partial x}
  \left( -\frac{\partial V(x,t)}{\partial x} P(x,t) \right)
  + \epsilon \frac{\partial^2 P(x,t)}{\partial x^2}, \label{FPeq}
\end{eqnarray}
where $P(x,t)$ and $J(x,t)$ are a probability density and
a probability current respectively.
We apply the periodic boundary conditions
on $P(x,t)$ and $J(x,t)$, 
\begin{eqnarray}
 P(x+L,t) &=& P(x,t), \label{P:spac_p}\\
 J(x+L,t) &=& J(x,t), \label{J:spac_p}
\end{eqnarray} 
where $P(x,t)$ is normalized in the spatial period $L$.
Except for transient time, $P(x,t)$ and $J(x,t)$ 
satisfy the temporally periodic conditions
\begin{eqnarray}
  P(x,t+T) &=& P(x,t),\label{P:temp_p}\\
  J(x,t+T) &=& J(x,t).\label{J:temp_p}
\end{eqnarray}

%----- energetics
According to the stochastic energetics \cite{Sekimoto-Energ}, 
the heat $\widetilde{Q}$ released to the heat bath during 
the period $T$ is given as,
\begin{equation}
 \widetilde{Q} =
  \int_{x(0)}^{x(T)} 
  \left\{-\left(-\frac{dx(t)}{dt}+\xi(t)\right)\right\} dx(t).
  \label{Def:til_Q}
\end{equation}
Inserting Eq. (\ref{LangevinEq}) into Eq. (\ref{Def:til_Q}),
we obtain the energy balance equation,
\begin{equation}
  \widetilde{Q} = \int_{0}^{T} \frac{\partial V(x(t),t)}{\partial t} dt
  - \int_{V(0)}^{V(T)} d V(x(t),t).
  \label{EnergyBalance_1}
\end{equation}
The first term of RHS is the energy $\widetilde{E_{in}}$ that the system obtain
through the potential modulation,
and the second term,
$\int_{V(0)}^{V(T)} d V(x(t),t)$,
is the work $\widetilde{W}$ 
that the system extracts from the input energy $\widetilde{E_{in}}$,
during the period $T$.
The ensemble average of $\widetilde{W}$ is given
using Eqs. (\ref{OR:potential}), (\ref{OR:temp_p}) and (\ref{P:temp_p})
as,
\begin{eqnarray}
 \left\langle \widetilde{W} \right\rangle 
  &=& \left\langle \int_{V(0)}^{V(T)} d V(x(t),t) \right\rangle \\ \nonumber
  &=& \ell \int_{0}^{T}dt \int^L_0 dx J(x,t) \equiv W,
  \label{OR:work}
\end{eqnarray}
where one can find that
 $W$ represents the work against the load.
Also, 
using Eqs. (\ref{OR:potential}), (\ref{FPeq})
and the periodic conditions (Eqs. (\ref{OR:temp_p}),
(\ref{OR:spac_p}), (\ref{J:spac_p}) and (\ref{P:temp_p})),
the ensemble average of $E_{in}$ is given as
\begin{eqnarray}
 \left\langle \widetilde{E_{in}} \right\rangle
  &=& \left\langle\int_{0}^{T} \frac{\partial V(x(t),t)}{\partial t} dt \right\rangle \\ \nonumber
  &=& \int_{0}^{T} dt\int^L_0 dx \left( -\frac{\partial V_{p}(x,t)}{\partial x} \right) J(x,t)
  \equiv E_{in}.
  \label{OR:inputenergy}
\end{eqnarray}
Taking an ensemble average, Eq. (\ref{EnergyBalance_1}) yields,
\begin{eqnarray}
 Q &=& E_{in} - W,  \label{EnergyBalance3}\\
  &=& \int_{0}^{T} dt \int^L_0 dx 
 \left( - \frac{\partial V_{p}(x,t)}{\partial x} \right) J(x,t)\\
 && - \ell \int_{0}^{T} dt \int^L_0 dx J(x,t),
  \label{EnergyBalance2}
\end{eqnarray}
where $Q\equiv  \left\langle \widetilde{Q} \right\rangle$.
Therefore we obtain the efficiency $\eta$ of the energy conversion
from the input energy $E_{in}$ into the work $W$, as follows,
\begin{equation}
 \eta = \frac{W}{E_{in}}
  = \frac{\ell \int_{0}^{T} dt \int^L_0 dx J(x,t)}
  {\int_{0}^{T}dt \int^L_0 dx 
  \left( -\frac{\partial V_{p}(x,t)}{\partial x} \right)J(x,t)}.
  \label{OR:efficiency}
\end{equation}
This expression can be estimated simply
by solving the Fokker-Planck equation (Eq. (\ref{FPeq})).

%----- Simulation
We solve Eq. (\ref{FPeq}) numerically
with the following ratchet potential as an example.
It satisfies Eqs. (\ref{OR:temp_p}), (\ref{OR:spac_p})
and the condition that
the degree of the asymmetry oscillates
but the amplitude of the ratchet is constant.
It will turn out that 
the results does not depend on 
the detailed shape of the potential.
The ratchet potential is 
\begin{equation}
 V_{p}(x,t)=\frac{1}{2} V_0
 \left( \sin \left(\frac{2 \pi x}{L} 
 + A(t) \sin \left(\frac{2 \pi x}{L} 
 + C_1\sin\left(\frac{2 \pi x}{L}\right) \right)\right) 
 + 1 \right),
 \label{OR:sim:potential}
\end{equation}
where
$A(t) = C_{2}+C_{3} \sin (\omega t)$,
and $V_0$, $C_1$, $C_2$, $C_3$ are constant.

%-----RESULT----
The results are shown in Fig. \ref{Fig:OR:sim_eff}.
We find that 
the efficiency is maximized at finite intensity of thermal noise 
(Fig. \ref{Fig:OR:sim_eff}(a)).
This shows that 
thermal noise can certainly facilitate 
the energy conversion.
What is the reason for the behavior of the efficiency $\eta$?
Let us see the work $W$ and the input energy $E_{in}$
as a function of the intensity of thermal noise.
The work $W$, the numerator of Eq. (\ref{OR:efficiency}),
has a peak at finite intensity of thermal noise
(Fig. \ref{Fig:OR:sim_W}(b)),
because of the behavior of the flow during the period $T$, 
$\bar{J}\equiv\int_{0}^{T}dt \int_{0}^{L}dx J$.
In the absence of thermal noise ($\epsilon=0$),
the particle cannot move over the potential peak
(which results in  $\bar{J}=0$).
As the intensity of thermal noise increases,
the effect of non-equilibility emerges
and it induces finite asymmetric flow against the load
through the asymmetry of the ratchet.
When thermal noise is large enough ($\epsilon\rightarrow\infty$),
the flow against load is no longer positive,
because the effect of the ratchet disappears in this limit.
Therefore the flow, and also the work, 
behave like Fig. \ref{Fig:OR:sim_W}(b)
as a function of thermal noise intensity.
The input energy $E_{in}$, the denominator of
Eq. (\ref{OR:efficiency}),
remains finite
at the limit $\epsilon\rightarrow0$ (Fig. \ref{Fig:OR:sim_E}(c)),
where
all input energy dissipates
because the oscillation of the local potential minimum makes
finite local current
even in the absence of thermal noise.
Therefore the efficiency starts with $\eta=0$
at $\epsilon=0$,
and grows up as the intensity of thermal noise increases,
then disappears as $\epsilon\rightarrow\infty$.
The efficiency has its peak at finite $\epsilon$.

%%%%%%%%%%%
As we have stated above, 
noise-induced flow and 
finite dissipation in the absence of thermal noise 
are the cause for the noise-induced energy conversion.
Thus our finding will not depend on the detail of the shape of $V_{p}(x,t)$.
We expect that thermal noise can facilitate the energy conversion
in a variety of  ratchet systems.

%===== forced thermal ratchet =====
Finally we discuss the forced thermal ratchet\cite{Magnasco-Force}.
The forced thermal ratchet is a system 
where a dissipative particle in a ratchet 
is subjected both to zero-mean external force
and to thermal noise.
The previous Letter\cite{Kamegawa_ea-Energ} 
was the first trial that discussed the energetics 
in the ratchet.
For the analytical estimate,
the discussion in that Letter
was only on the quasi-static limit
where the change of the external force is slow enough.
In that case, thermal noise cannot facilitate operation of the ratchet. 
The energetic efficiency
is monotonically decreasing function
of thermal noise intensity,
in contrast to the oscillating ratchet discussed above.
However
one notices that 
the external force of the forced thermal ratchet
can also be written by oscillatory modulating potential,
when the external force is periodic
as in the literature\cite{Magnasco-Force,Kamegawa_ea-Energ},
It is likely that 
the difference 
between the two cases,
the oscillating ratchet
and the forced thermal ratchet discussed 
in that Letter\cite{Kamegawa_ea-Energ},
is attributed to
the condition of the system,
namely, quasi-static or not.
We suppose that
thermal noise may facilitate 
the energy conversion 
in the forced thermal ratchet
when the ratchet is {\it not\/} quasi-static.

Langevin equation of the forced thermal ratchet 
is the same as Eq. (\ref{LangevinEq}),
except for the potential $V$.
In this case, the potential is 
\begin{equation}
 V(x,t)=V_{p}(x) + \ell x - F_{ex}(t) x, 
  \label{FTR:potential}
\end{equation}
where 
$V_{p}(x)$, $\ell$ and $F_{ex}$ represent
the ratchet potential, load and 
an external force respectively.
The periodic external force $F_{ex}(t)$ satisfies 
$F_{ex}(t+T)=F_{ex}(t)$ and 
$\int_{0}^{T}dt F_{ex}(t)=0$\cite{fn:Fex}.
The work $W$ is the same as Eq. (\ref{OR:work}),
and the input energy $E_{in}$ is,
\begin{equation}
 E_{in} = \int_{0}^{T}dt\int_{0}^{L}dx F_{ex}(t) J(x,t).
\end{equation}
In quasi-static limit \cite{Kamegawa_ea-Energ}, 
the probability current $J$ does not depend on the coordinate $x$.
Thus, when the current over the potential peak 
(that causes $W$)
vanishes,
the local current 
vanishes anywhere ($J(x,t)=J(t)=0$).
However, if the system is not quasi-static,
the behavior changes qualitatively.
In this case,
even when the current over the potential peak vanishes at $\epsilon=0$,
local current around the local potential minimum still remains finite.
Thus there exists finite energy dissipation even in the limit
$\epsilon\rightarrow0$, 
which means that 
the input energy $E_{in}$ still remains finite value 
at this limit (Fig. \ref{Fig:FTR:sim_eff}(c)).
Therefore, the efficiency is found to be zero at $\epsilon=0$,
and has a peak at finite $\epsilon$ (Fig. \ref{Fig:FTR:sim_eff}(a)).
The result is the same as that of the oscillating ratchet.
It must be noted that the energetics can distinguish
the behavior of the efficiency
in the non-quasistatic case from that in quasi-static case,
although the dependences of the flow $\bar{J}$ are 
the same between the two.

%=====conclusion========
We have discussed energetics of the ratchet system
using the method of the stochastic energetics,
and estimated the efficiency of energy conversion.
We found that
thermal noise {\it can} facilitate the operation of the ratchet system.
%*****
The mechanism was briefly summarized as follows:
Through the ratchet,
potential modulation 
causes noise-induced flow against the load
that results in the work.
On the other hand,
potential modulation with finite speed causes
local current around the local potential minimum that makes
finite dissipation even in the absence of thermal noise.
Thus
the efficiency is maximized at finite intensity of thermal noise.
%*****
The result must be robust and independent of the detail 
of the potential,
because only two factors are essential
for the energy conversion activated by thermal noise:
One is the noise-induced flow,
and the other is 
the finite dissipation in the absence of thermal noise.
Also in the two-state model\cite{Julicher-Model}
that is an other type of ratchet systems,
it was reported quite recently that
the efficiency could be maximized at finite temperature\cite{Prost}.
We expect it to be examined by experiment whether and how 
the real molecular motors use thermal noise.

%===== Acknowledgment =====
We would like to thank K. Sekimoto, 
J. Prost, A. Parmeggiani, F. J\"{u}licher, S. Sasa,
T. Fujieda and T. Tsuzuki for helpful comments.
This work is supported by the Japanese Grant-in-Aid for Science Research
Fund from the Ministry of Education, Science and Culture (No. 09740301) and
Inoue Foundation for Science.

%========================================
%     References
%========================================

%===== figures =====
\begin{figure}
 \caption{Oscillating ratchet potential $V_{p}(x,t)$.
 The ratchet potential changes continuously 
 between solid line and broken line with the time period $T$.
 The amplitude of the ratchet keeps constant, $V_{0}$.}
 \label{Fig:OscillatingPotential}
\end{figure}
\begin{figure}
 \caption{Snapshot of the potential $V(x,t)$ (solid line).
 Broken line represents the load term, $\ell x$.
 }
 \label{Fig:OR:loaded}
\end{figure}
\begin{figure}
 \caption{The energetic efficiency, $\eta=W/E_{in}$,
 of the oscillating ratchet system 
 as a function of thermal noise intensity,
 where $V_0/\omega L^2=0.01$, $\ell/\omega L=0.00002$,
 $C_1=0.3$, $C_2=0.3$ and $C_3=0.3$.
 (a) the efficiency $\eta$,
 (b) the work  $W$ and (c) the input energy $E_{in}$.}
 \label{Fig:OR:sim_eff}
 \label{Fig:OR:sim_W}
 \label{Fig:OR:sim_E}
\end{figure}
\begin{figure}
 \caption{The energetic efficiency, $\eta=W/E_{in}$,
 of the forced thermal ratchet
 as a function of thermal noise intensity,
 where $V_0/\omega L^2=1.0$, $\ell/\omega L=0.001$
 and $\left|F_{ex}\right|_{max}/\omega L=1.0$.
 (a) the efficiency $\eta$,
 (b) the work  $W$ and (c) the input energy $E_{in}$.}
 \label{Fig:FTR:sim_eff}
\end{figure}


\begin{references}
%%%%% Experiences %%%%%
 \bibitem{Yanagida_ea-Slidi}
 % Sliding distance of actin filament induced by a myosin crossbridge
 % during one ATP hydrolysis cycle
 T. Yanagida, T. Arata, and F. Oosawa, Nature {\bf 316}, 366 (1985).

 \bibitem{Ishijima-Multi}
 % Multiple- and single-Molecule analysis of the actomyosin motor 
 % by nanometer-piconewton manipulation with a microneedle:
 % unitary steps and forces
 A. Ishijima, H. Kojima, H. Higuchi, Y. Harada, T. Funatsu, and T. Yanagida,
 Biophys. J. {\bf 70}, 383 (1996).

 \bibitem{Uyeda_ea-Myosi}
 % Myosin step size 
 % Estimation from slow sliding movement of actin over low densities of
 % heavy meromyosin
 T. Q. P. Uyeda, S. J. Kron, and J. A. Spudich,
 J. Mol. Biol. {\bf 214}, 699 (1990).

 \bibitem{Yasuda_ea}
 R. Yasuda, H. Noji, K. Kinosita, F. Motojima, and M. Yoshida,
 J. Bioenerg. Biomembr. {\bf 29}, 207 (1997).

%%%%% Vale-Oosawa %%%%%
 \bibitem{Vale_ea-Prote}
 R. D. Vale and F. Oosawa,
 Adv. Biophys., {\bf 26,} 97 (1990).

%%%%% A. F. Huxley %%%%%
 \bibitem{AHuxley}
 A. F. Huxley and R. M. Simmons,
 Nature, {\bf 233}, 533 (1971).

%%%%% Prost Review %%%%%
 \bibitem{Julicher-Model}
 See, e.g.,
 F. J\"{u}licher, A. Ajdari and J. Prost,
 Rev. Mod. Phys. {\bf 69} 1269 (1997),
 and references therein.

%%%%% Magnasco FTR %%%%%
 \bibitem{Magnasco-Force}
 M. O. Magnasco,
 Phys. Rev. Lett. {\bf 71,} 1477 (1993). 

%%%%% energetics FTR %%%%%
 \bibitem{Kamegawa_ea-Energ}
 H. Kamegawa, T. Hondou and F. Takagi,
 Phys. Rev. Lett., {\bf 80,} 5251 (1998).

%%%%% Stochastic Energetics %%%%%
 \bibitem{Sekimoto-Energ}
 K. Sekimoto, 
 J. Phys. Soc. Jpn. {\bf 66,} 1234 (1997).

%%%%% Sekimoto Sasa %%%%%
\bibitem{Sekimoto_ea-Compl}
 K. Sekimoto and S. Sasa,
 J. Phys. Soc. Jpn., {\bf 66,} 3326 (1997).

%%%%% Miki %%%%%
 \bibitem{Matsuo-Stoch}
 M. Matsuo and  S. Sasa,
 Physica A (to be published).

%%%%% 
 \bibitem{Hondou_ea-Irrev}
 T. Hondou and F. Takagi,
 J. Phys. Soc. Jpn., {\bf 67}, 2974, (1998)

%%%%%
 \bibitem{Sekimoto_ea-Molec}
 K. Sekimoto, F. Takagi and T. Hondou, cond-mat/9904322.

%%%%% Footnote %%%%%
\bibitem{fn:chemical}
In this paper, we discuss the systems which
convert mechanical energy into mechanical work,
while real molecular motors in biological systems
convert chemical energy into mechanical work.
Recently
an experiment suggests
that the protein can store 
the chemical energy from ATP hydrolysis\cite{Ishijima_ea-Simul},
the energy of which may be stored 
in mechanical way,
for example, by conformational change of the protein.
Some models have been proposed to explain this kind
of energy storage\cite{Nakagawa_et-prepri}.

%%%%% Risken %%%%%
 \bibitem{Risken_FPbook}
 H. Risken,
 {\it The Fokker-Planck Equation} 2nd ed., 
 (Springer-Verlag Berlin, 1989).

%%%%% footnote %%%%%
\bibitem{fn:Fex}
We consider the low amplitude regime\cite{Kamegawa_ea-Energ}
where the amplitude of $F_{ex}(t)$ is small.
In this case,
a particle cannot move over the potential peak
without thermal noise,
as in case of the oscillating ratchet.

%%%%% Prost %%%%%
 \bibitem{Prost}
 A. Parmeggiani, F. J\"{u}licher, A. Ajdari and J. Prost, 
 cond-mat/9904153.

%%%%% experiment %%%%%
 \bibitem{Ishijima_ea-Simul}
 A. Ishijima, H. Kojima, T. Funatsu, M. Tokunaga, H. Higuchi, 
 H. Tanaka, and T. Yanagida,
 Cell {\bf 92}, 161 (1998).

%%%%% Nakagawa %%%%%
 \bibitem{Nakagawa_et-prepri}
 For example, N. Nakagawa and K. Kaneko, chao-dyn/9903005.

%%%%% Feynman %%%%%
 \bibitem{Feynman_ea}
 R. P. Feynman, R. B. Leighton and M. Sands,
 {\em The Feynman Lectures in Physics} 
 (Addison-Wesley Publ. Co., Reading, Massachussets, 1966), vol. I.

\end{references}
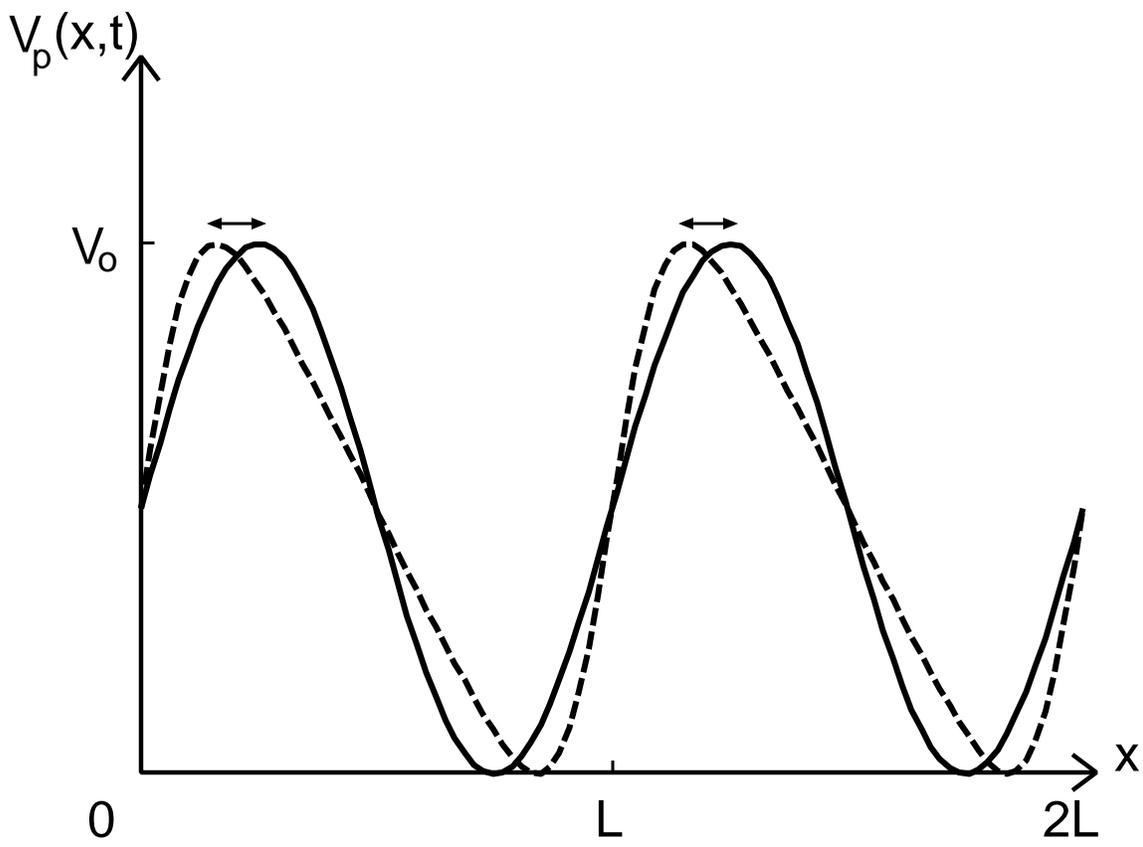
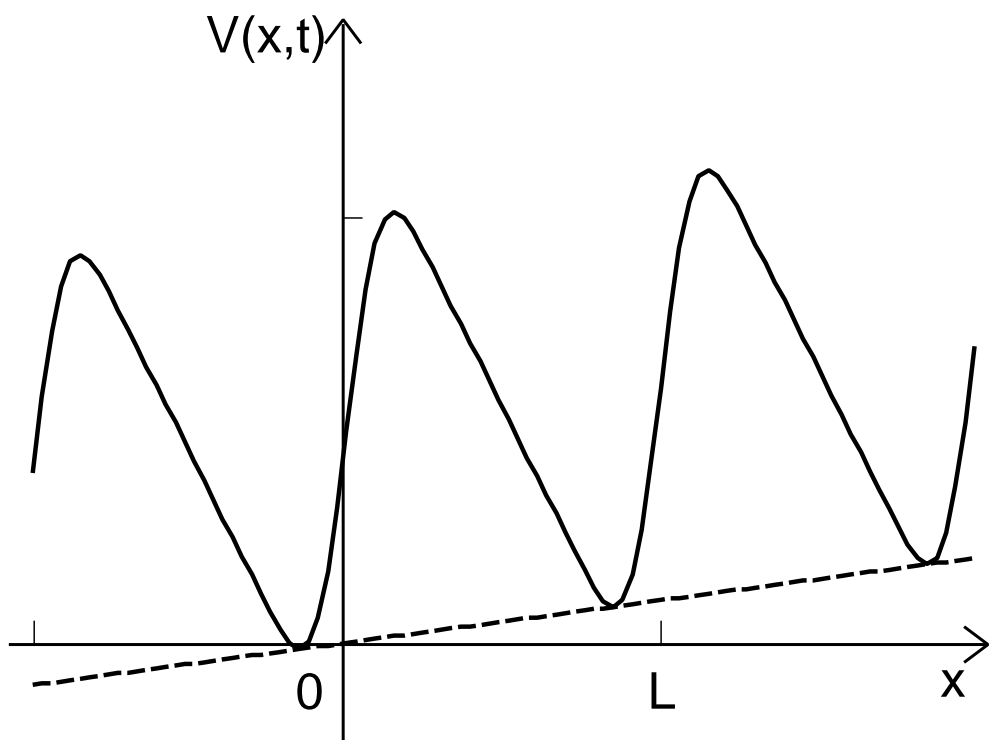
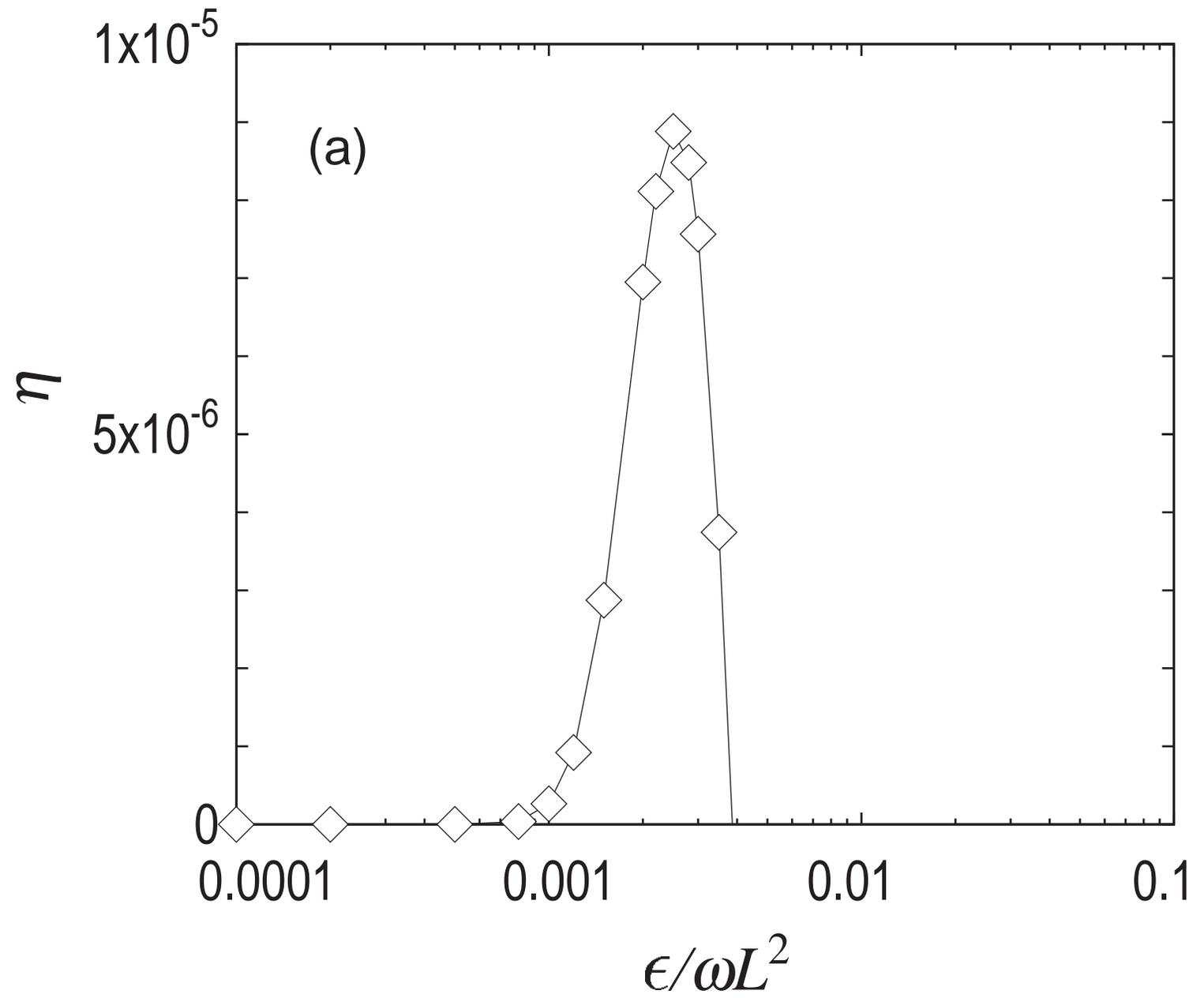
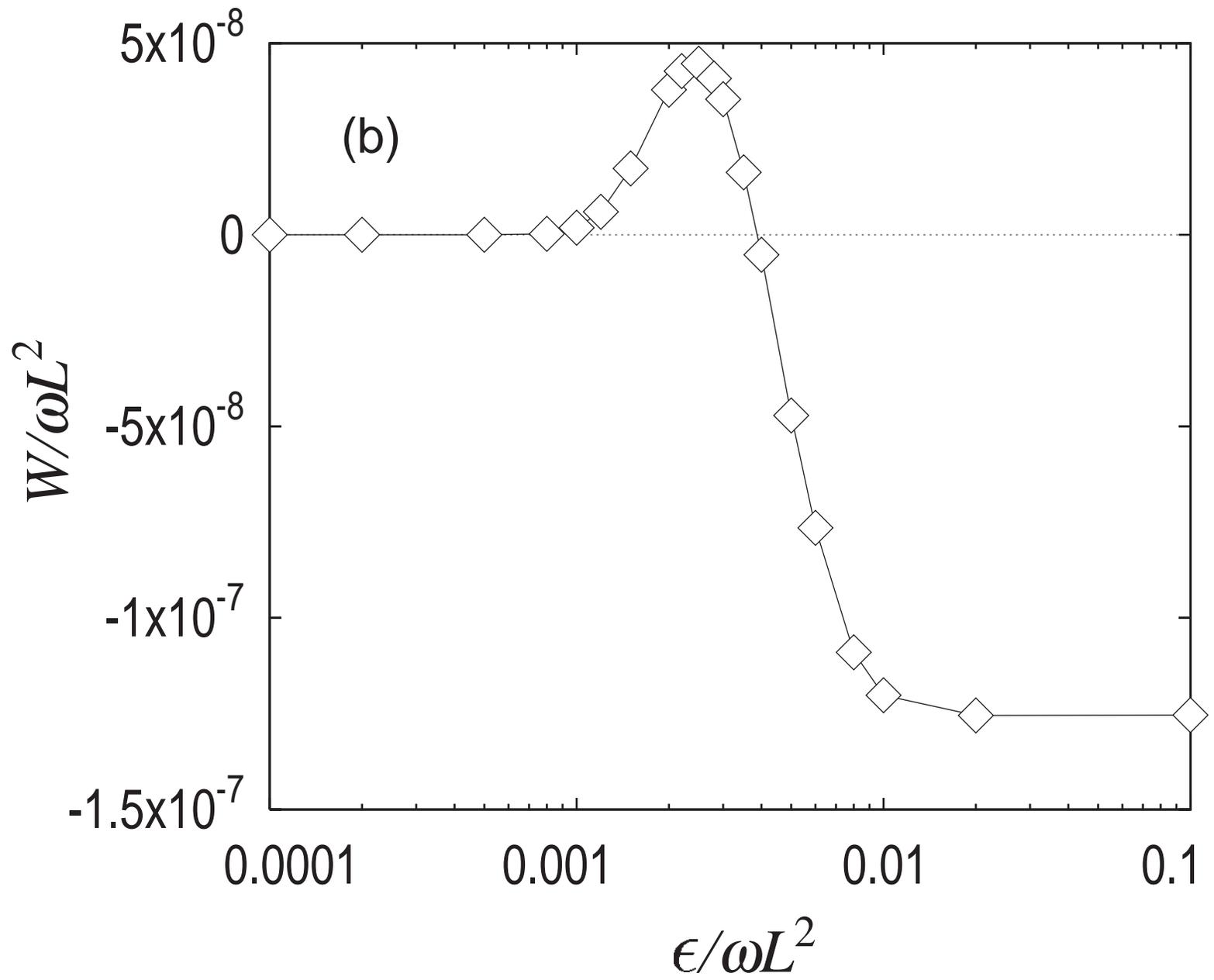
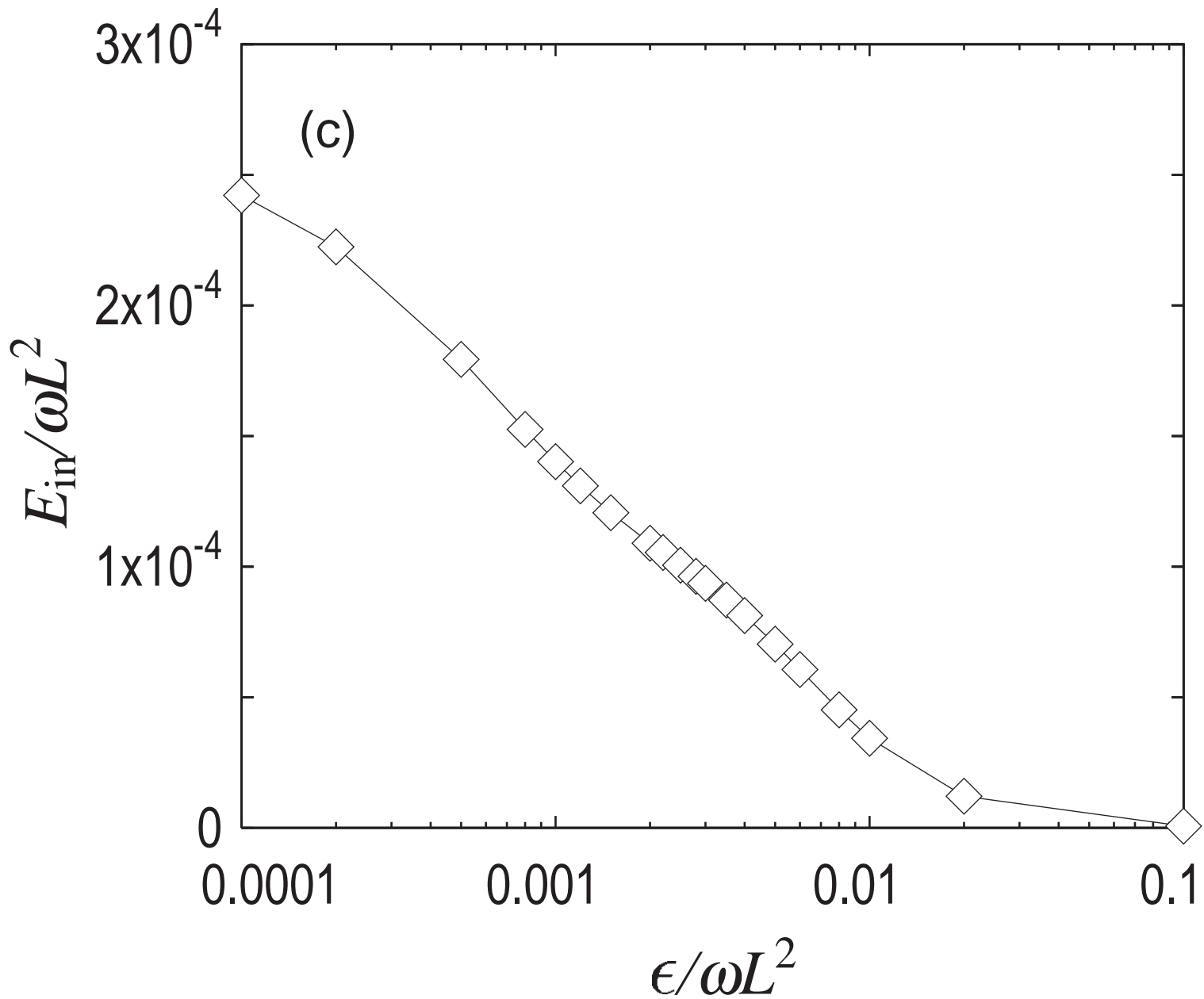
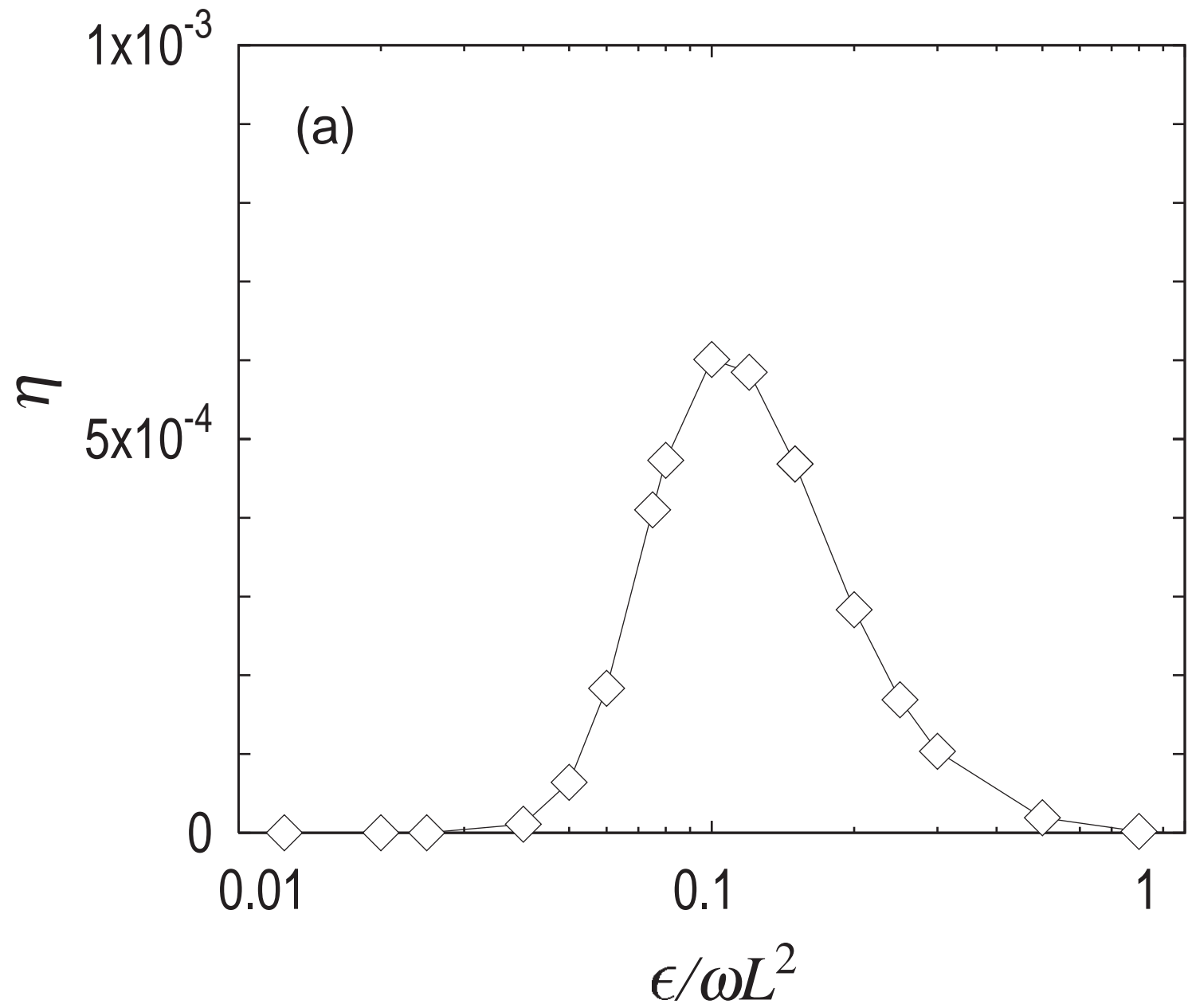
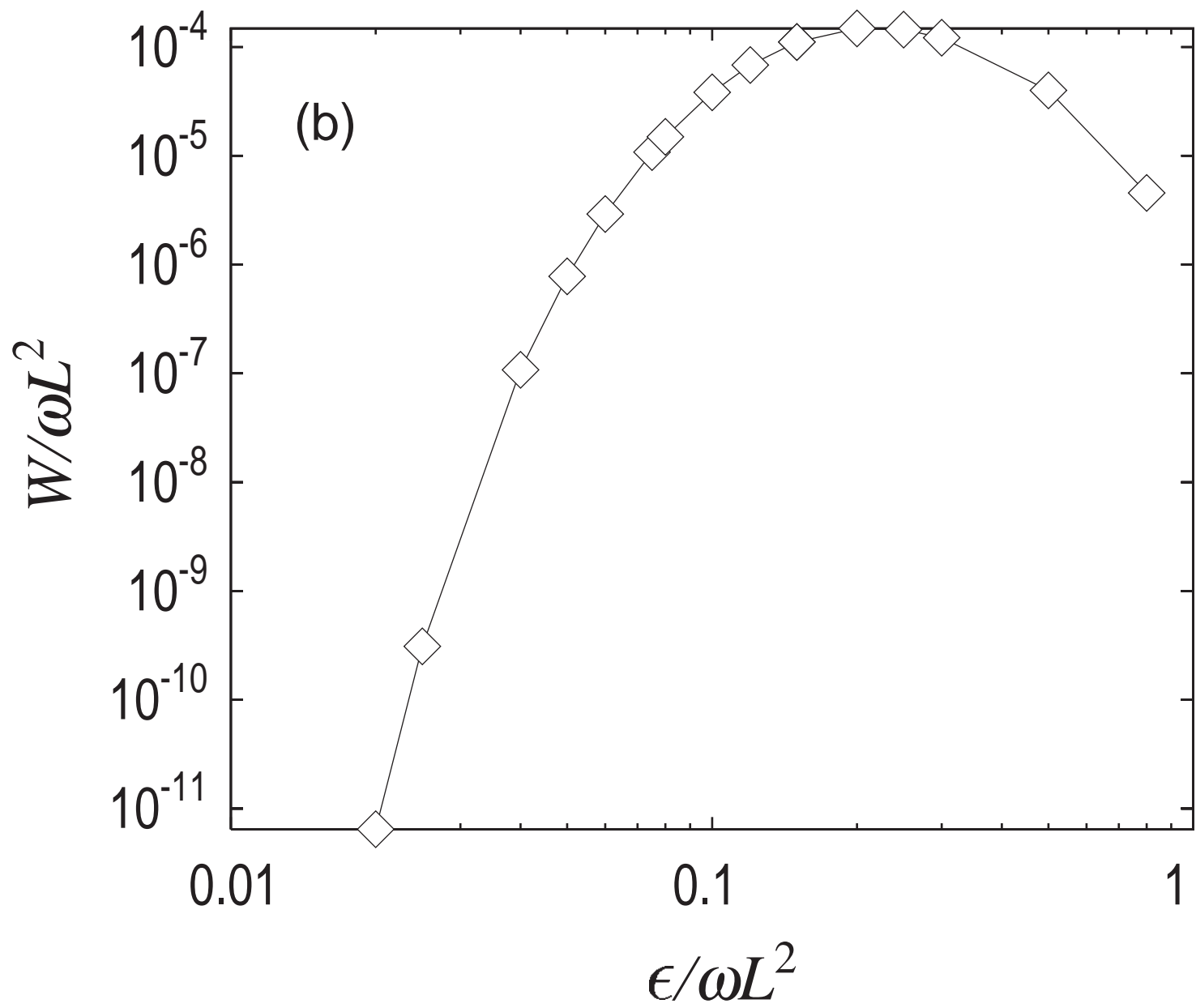
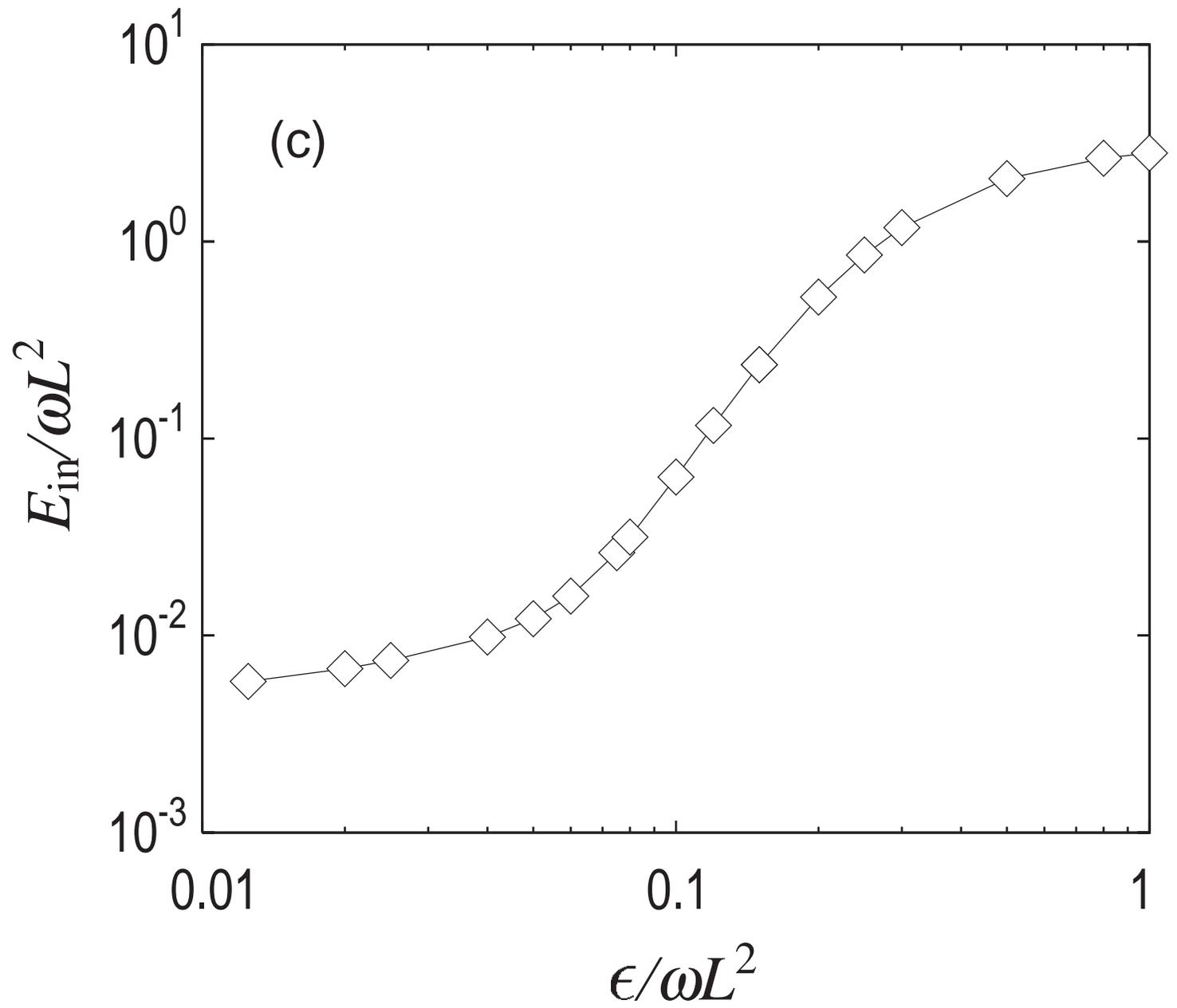
\end{document}